\documentstyle[prd,aps,floats]{revtex}

\begin{document}
\draft

\input epsf
\renewcommand{\topfraction}{0.99}
\twocolumn[\hsize\textwidth\columnwidth\hsize\csname 
@twocolumnfalse\endcsname

\title{Inflation on a single brane -- exact solutions}
\author{Rachael M. Hawkins and James E. Lidsey} 
\address{ Astronomy Unit, School of Mathematical Sciences,
Queen Mary, University of London,\\ Mile End Road, London, E1
4NS,~~U.~~K.  } 
\date{\today} \maketitle
\begin{abstract}
Algorithms are developed for 
generating a class of exact braneworld cosmologies, 
where a self--interacting scalar field is confined to a 
positive--tension brane embedded in a bulk containing a 
negative cosmological constant. It is assumed that the five--dimensional 
Planck scale exceeds the brane tension but 
is smaller than the four--dimensional Planck mass.
It is shown that 
the field equations can be expressed as a first--order system. 
A number of solutions to the equations 
of motion are found.  The potential resulting in the perfect fluid 
model is identified. 

\end{abstract}


\vskip2pc]

\section{Introduction}

\setcounter{equation}{0}

The possibility \cite{braneref,rs}
that our observable 
universe may be viewed as a domain wall embedded in a 
higher--dimensional space has received considerable 
attention recently. 
Motivated by developments 
in superstring and M--theory \cite{hw}, 
it is assumed in this scenario that 
the standard model interactions are confined 
to a $(3+1)$--dimensional hypersurface, but 
that gravity 
may propagate through the `bulk' dimensions 
perpendicular to the brane. 
This change in viewpoint has important consequences 
for early universe cosmology and, in particular, 
for the inflationary paradigm. 

Various cosmological aspects of branes 
embedded in five dimensions have been investigated 
\cite{bine,flanagan,4a,shiromizu,maartens,radterm,radterm1,easier,others,bulk,bulk1,projection}. 
The effects of including a scalar field in the bulk have also 
been studied
\cite{bulk,bulk1,projection}. 
In this paper, we consider 
a co--dimension $1$ brane with positive tension, $\lambda$, 
embedded in vacuum Einstein gravity with a negative cosmological 
constant. This corresponds to the scenario 
introduced by Randall and Sundrum where the extra 
dimension is infinite \cite{rs}. 
We focus on the region of parameter space 
defined by $\lambda^{1/4} \ll m_5 \ll m_4$, where 
$m_{4,5}$ represent the four-- and five--dimensional 
Planck scales, respectively.  

One approach to determining the 
expansion of the brane
is to project 
the five--dimensional 
metric onto the brane world--volume \cite{shiromizu,projection}.
In effect, this is equivalent to 
solving Einstein's field equations, 
together with appropriate jump conditions, for a
negative cosmological constant 
and an energy--momentum 
tensor restricted to the brane \cite{bine}.
The field equations admit a first integral that may be 
interpreted as a generalized Friedmann equation \cite{bine,flanagan,shiromizu}. 
We emphasize, however, that 
the dynamics could differ in a compactified scenario, due to 
the effects of radion stabilization.
If the cosmological constant and 
brane tension are related in an appropriate fashion, this equation 
takes the form\footnote{In 
this 
paper we assume 
that the  world--volume metric of the 
three--brane is the spatially 
flat, Friedmann--Robertson--Walker (FRW) line 
element with scale factor,  $a (t)$, and 
Hubble parameter, $H \equiv \dot{a}/a$. 
A dot and prime denote differentiation 
with respect to cosmic time, $t$, and the 
scalar field, $\phi$, respectively.} 
\cite{bine,flanagan}:
\begin{equation}
\label{Friedmann}
H^2 = \frac{4\pi}{3 \lambda m^2_4} \rho (
\rho +2 \lambda ) + \frac{\epsilon}{a^4}  , 
\end{equation}
where
$\rho$ represents the 
energy density of matter on the 
brane, $\epsilon$ is a constant and we have 
assumed that  the four--dimensional 
cosmological constant is zero. The last term on the 
right--hand side of Eq. (\ref{Friedmann}) behaves 
as `dark radiation' and arises due to the 
backreaction of the bulk gravitational degrees of freedom on the 
brane \cite{bine,flanagan,radterm,radterm1}. 
At sufficiently low energies, 
$\rho \ll \lambda$, the standard cosmic 
behaviour is recovered and  
the primordial nucleosynthesis constraint 
is satisfied provided that 
$\lambda \ge (1 \, {\rm MeV})^4$. 
 
Quantum gravitational effects become important if the energy density on 
the brane exceeds the five--dimensional Planck scale. 
In this case, the assumption that matter is confined 
to the brane may become unreliable. However,  
if $\lambda \ll m^4_5$, there is a region of parameter 
space, corresponding to $\lambda \ll \rho \ll m^4_5$, where 
the classical solution is still 
valid, 
but 
where the quadratic correction becomes 
significant.   
In particular, 
this
term can play an important role 
during inflation \cite{maartens,easier}. 
Cosmological inflation has played a central 
role in studies of the very early universe. (For 
recent reviews, see, e.g., Refs. \cite{infrev,infrev1}). 
It is therefore important to 
study the inflationary dynamics of the 
braneworld scenario. 
In the standard inflationary cosmology, 
the universe is dominated by  
a scalar `inflaton' field 
self--interacting through 
a potential, 
$V(\phi)$,  with 
an energy density $\rho \equiv \dot{\phi}^2/2 +V$. 
Inflation proceeds if the potential energy of the field 
dominates its kinetic energy. Maartens {\em et al.} 
have recently considered the case 
where a single inflaton field 
is confined to the brane and have 
derived the necessary criteria for successful inflation
when the slow--roll approximation is valid \cite{maartens}.
In general, 
the quadratic term in the Friedmann equation 
(\ref{Friedmann}) results in an enhanced friction 
on the field, implying that inflation 
is possible for a wider region of parameter 
space than in the standard cosmology 
\cite{4a,maartens,radterm1,easier}.

When the scalar field is confined to the 
brane, 
energy--momentum  conservation 
implies that its equation 
of motion has the standard form: 
\begin{equation}
\label{phieom}
\ddot{\phi}+3H \dot{\phi} + V'(\phi) =0   . 
\end{equation}
Eq. (\ref{phieom})  can be expressed in terms of the 
energy density such that 
\begin{equation}
\label{scalareom}
\dot{\rho} = -3H\dot{\phi}^2
\end{equation}
and it follows that a given inflationary braneworld 
model is specified 
by a solution to the Friedmann--scalar 
field equations
(\ref{Friedmann}) and (\ref{scalareom}). 

In general, however, it is very difficult 
to solve this system of equations,  even 
when the 
standard slow--roll assumptions, 
$\dot{\phi}^2 \ll V$ and $|\ddot{\phi} | \ll 
H |\dot{\phi} |$,  are made. 
It is therefore important to develop 
generating techniques 
for finding exact solutions 
to Eqs. 
(\ref{Friedmann}) and (\ref{phieom}) and 
this is the purpose of the 
present work. 
During inflation, the dark radiation redshifts rapidly 
and soon becomes dynamically negligible
and we therefore consider models where 
$\epsilon =0$. On the other hand, 
we do not assume that the slow--roll 
approximation 
is necessarily valid. Thus, such a study has 
direct applications to the 
final stages of inflation, 
where the kinetic energy of the inflaton 
field 
inevitably becomes 
significant. Depending on the 
form of the potential, this may happen when the quadratic correction 
to the Friedmann equation 
(\ref{Friedmann}) is still important. 
The search for exact scalar field braneworlds 
is also important because it allows one to classify the 
possible types of behaviour that arise in these universes and 
to uncover the generic characteristics of 
such models. 

\section{First--Order Field Equations}

\setcounter{equation}{0}

From a particle physics perspective, it 
is natural to begin by specifying the 
functional form of the potential. 
However, even for 
simple choices, such as exponential or power law 
potentials, 
analytical progress can not be made. 
An alternative route, 
originally introduced 
within the context of standard 
chaotic inflation \cite{startscale}, 
is to specify the 
time--dependence of the scale factor, $a(t)$. In the braneworld 
scenario, given $a(t)$, one may deduce $\rho(t)$ from Eq. 
(\ref{Friedmann}) and hence $\dot{\phi}(t)$ from 
Eq. (\ref{scalareom}). Integrating yields 
$\phi (t)$. 
The 
time--dependence of the potential follows immediately from 
the definition of the energy 
density and the form of $V(\phi)$ is deduced 
by inverting $\phi (t)$. The drawback of this
approach is that  such an inversion is not always 
possible and moreover a realistic 
potential does not 
necessarily result. 

In principle, 
these problems are partially avoided by first specifying 
an invertible form for $\phi (t)$
\cite{barrowparsons}, although 
we do not explore this possibility further here. 
Our approach is to note that during inflation, 
the scalar 
field rolls monotonically down its potential. 
Thus, 
Eq. (\ref{scalareom}) may be expressed in the 
form 
\begin{equation}
\label{first}
\rho' =-3H\dot{\phi}    . 
\end{equation}
The formal limit,  
$\lambda \rightarrow \infty$, corresponds to 
the standard 
inflationary scenario. In this case, 
Eqs. (\ref{Friedmann}) and (\ref{first}) 
can be written 
in the particularly 
simple `Hamilton--Jacobi' form \cite{hj,lidsey}
\begin{eqnarray}
\label{fieldlimit1}
H'a' = -\frac{4\pi}{m_4^2} Ha \\
\label{fieldlimit2}
H' =-\frac{4\pi}{m_4^2} \dot{\phi} \\
\label{fieldlimit3}
(H')^2- \frac{12 \pi}{m^2_4}H^2 = -\frac{32\pi^2}{m^4_4}V  , 
\end{eqnarray}
where 
the Hubble parameter is viewed implicitly as a function 
of the scalar field.  
Eq. (\ref{fieldlimit1}) may be integrated 
to yield the scale factor: 
\begin{equation}
\label{standardscale}
a (\phi) =\exp \left[ -\frac{4\pi}{m^2_4} \int^{\phi}d\phi 
\frac{H}{H'} \right]
\end{equation}
and this implies that 
the cosmological dynamics 
is determined, 
up to a single quadrature,
once the functional form of $H(\phi )$ has been specified. 
It has been suggested 
that $H(\phi )$ should 
be viewed as the solution generating function 
when analysing inflationary cosmologies \cite{lidsey1}. The advantage 
of such an approach 
is that the form of the potential is readily deduced from Eq. 
(\ref{fieldlimit3}).
The question that naturally arises within the braneworld 
context, therefore, is whether there exists an analogous 
generating 
function when the quadratic correction in the Friedmann equation 
(\ref{Friedmann}) is relevant.

To proceed, we 
define a new function $y(\phi )$:
\begin{equation}
\label{fdefine}
\rho \equiv \frac{2\lambda y^2}{1-y^2}  , 
\end{equation}
where the restriction $y^2<1$ must be 
imposed for the 
weak energy condition to be satisfied. Substituting 
Eq. (\ref{fdefine}) into the Friedmann equation
(\ref{Friedmann}) implies that the Hubble parameter 
is given by
\begin{equation}
\label{fHubble}
H (\phi ) = \left( \frac{16\pi \lambda}{3m^2_4} 
\right)^{1/2} \frac{y}{1-y^2}  . 
\end{equation}
The 
right--hand side of Eq. (\ref{fHubble}) 
can be expanded as a geometric progression: 
\begin{equation}
\label{geometric}
H = \left( \frac{16\pi \lambda}{3m^2_4} \right)^{1/2}
\left[ y +y^3+y^5 + \ldots \right]     , 
\end{equation}
implying  that 
$y$ is proportional to the Hubble 
parameter
in the low--energy limit, 
$y 
\rightarrow 0$ $(\rho/ \lambda \rightarrow 
0 )$.

Substitution of Eq. (\ref{fdefine}) 
into the scalar field equation (\ref{first}) 
implies that
\begin{equation}
\label{fprime}
\dot{\phi} = - \left( \frac{\lambda m^2_4}{3\pi} \right)^{1/2}
\frac{y'}{1-y^2} 
\end{equation}
and the potential is given in terms of $y(\phi )$ 
by combining Eqs. (\ref{fdefine}) and (\ref{fprime}):
\begin{equation}
\label{fpotential}
V (\phi ) = \frac{2\lambda y^2}{1 -y^2}  
- \frac{\lambda m^2_4}{6 \pi} \left( \frac{y'}{1-y^2} 
\right)^2   . 
\end{equation}
It follows from Eqs. (\ref{fHubble}) and (\ref{fprime}) 
that the scale factor satisfies 
\begin{equation}
\label{fsatisfy}
y'a' =-\frac{4\pi}{m^2_4} ya  . 
\end{equation}
Eq. (\ref{fsatisfy}) is formally 
identical to Eq. (\ref{fieldlimit1}) 
and 
may be integrated to yield the scale factor 
in terms of a 
single quadrature with respect to the scalar field:
\begin{equation}
\label{fscale}
a(\phi ) = \exp \left[ -\frac{4 \pi}{m^2_4} 
\int^{\phi} d \phi \frac{y}{y'}
\right]  . 
\end{equation}
Finally, the dependence 
of the scalar 
field on cosmic time is deduced by 
evaluating the integral 
\begin{equation}
\label{tf}
t-t_0 = \left( \frac{3\pi}{\lambda m^2_4}
\right)^{1/2} \int^{\phi}_{\phi_0} d\phi \frac{y^2 -1}{y'} 
\end{equation}
and inverting the result, where $t_0$ is an arbitrary integration 
constant.

Thus, we have reduced the 
second--order system of 
equations (\ref{Friedmann}) and (\ref{phieom}) 
to the non--linear,  
first--order system (\ref{fprime})
and (\ref{fsatisfy}) by employing 
the scalar field as dynamical variable.  

Further insight may be gained 
by defining a second new function $b(\phi)$:
\begin{equation}
\label{definitionb}
y \equiv {\rm tanh} b   . 
\end{equation}
It follows immediately from Eqs. (\ref{fHubble}) and 
(\ref{fprime}) that
the Hubble parameter is given by
\begin{equation}
\label{bHubble}
H=\left( \frac{4\pi \lambda}{3m^2_4} \right)^{1/2}
\sinh 2b
\end{equation}
and that the scalar field varies as 
\begin{equation}
\label{bkinetic}
\dot{\phi} =-\left( \frac{\lambda m^2_4}{3\pi} 
\right)^{1/2} b'  . 
\end{equation}
Eq. (\ref{bkinetic}) is formally 
equivalent to Eq. (\ref{fieldlimit2}). 
Hence, the scale factor is given by 
\begin{equation}
\label{bscale}
a=\exp \left[ -\frac{2\pi}{m^2_4} \int^{\phi} 
d\phi \frac{\sinh 2b}{b'} \right]
\end{equation}
and the potential takes the simple form 
\begin{equation}
\label{bpotential}
V= 2\lambda \sinh^2 b -\frac{\lambda m^2_4}{6\pi} b'^2  . 
\end{equation}
The time--dependence of the scalar field is determined by
the integral 
\begin{equation}
\label{btime}
t-t_0 = -\left( \frac{3\pi}{\lambda m^2_4} \right)^{1/2}
\int^{\phi}_{\phi_0} d\phi \frac{1}{b'}  .
\end{equation}
The function $b$ plays the equivalent role 
to that of the Hubble parameter 
in the field equation (\ref{fieldlimit2}). Indeed, 
$b (\phi)  \propto H (\phi)$ in the low--energy
limit. It follows from Eq. (\ref{bkinetic}) that 
$b (t)$ is a monotonically decreasing function of 
cosmic time. 

To summarize this Section, we 
have found that the braneworld field equations 
can be rewritten after suitable redefinitions 
in a way that directly extends
the Hamilton--Jacobi form of the 
standard 
scalar field 
cosmology. 
However, the 
physical interpretation of the variables 
is different in the two cases. 
Nevertheless, this 
correspondence implies that 
similar techniques may be employed to find exact 
braneworld cosmologies. In particular, 
when  the functional form of the 
parameter $y(\phi)$ is known, the potential 
is determined in terms of this function and 
its first derivative. The function 
$a(\phi)$ follows from Eq. (\ref{fscale}) and 
$\phi (t)$ follows 
by evaluating Eq. (\ref{tf}) 
and inverting the result. 
An alternative method for 
solving the field equations 
is to
specify the form of $b(\phi)$ and 
to then evaluate 
the two integrals in Eqs. (\ref{bscale}) and 
(\ref{btime}).
This is equivalent to 
determining the kinetic energy of the 
scalar field as a function of the field itself. 
A related technique was recently employed in the standard inflationary 
scenario \cite{relatedkinetic}.
The advantage of this approach is that the integrand in 
Eq. (\ref{bscale}) can be expanded as a power 
series in $b$. 

\section{Exact Braneworlds}

\setcounter{equation}{0}

We now find exact braneworld models 
by employing the above techniques. 
Firstly, we consider the ansatz
\begin{equation}
\label{ansatz1}
y ={\rm sech}  \left( \frac{\sqrt{2\pi}C}{m_4} \phi \right) , 
\end{equation}
where $C$ is an arbitrary constant. 
Substitution of Eq. (\ref{ansatz1}) into Eq. 
(\ref{fpotential}) implies that the potential 
is given by 
\begin{equation}
\label{fluidpotential}
V = \frac{\lambda}{3} \left( 6-C^2 \right)
{\rm cosech}^2  \left( \frac{\sqrt{2\pi}C}{m_4}\phi \right)  . 
\end{equation}
Thus, we require $C^2 <6$ for the 
potential to be positive--definite and 
we consider this region of parameter 
space in what follows. 
Evaluating Eq. (\ref{tf}) implies that 
\begin{equation}
\label{tfevaluate}
t-t_0 = \left( \frac{3}{4\pi \lambda} \right)^{1/2} \frac{m_4}{C^2}
\cosh 
\left( \frac{\sqrt{2\pi} C}{m_4} \phi \right) 
\end{equation}
and substituting Eqs. (\ref{ansatz1}) and (\ref{tfevaluate})
into Eq. (\ref{fscale}) implies that the scale factor is given by
\begin{equation}
\label{scaletime}
a(t) = \left[ \frac{4\pi\lambda C^4}{3m^2_4}
\left( t-t_0 \right)^2 
-1 \right]^{1/C^2}  . 
\end{equation}
Without loss of generality, we may  choose 
$t_0 = - [ 3m^2_4/(4\pi \lambda C^4) ]^{1/2}$ 
such that the
origin of time corresponds to a vanishing scale factor.

It can be verified by direct substitution that 
the solution  (\ref{scaletime}) is a scaling 
solution, in the sense that the kinetic and potential 
energies of the scalar field redshift at the same 
rate as the brane expands. The field behaves 
as a perfect fluid with an 
effective equation of state, $p =\omega \rho$, 
where the barotropic index is given by 
\begin{equation}
\label{barotropic}
\omega = \frac{C^2-3}{3}   . 
\end{equation}
Thus, we have found the scalar field model 
equivalent to the perfect fluid cosmology presented in  
Ref. \cite{bine}.  
The solution is interesting because 
it reduces to the power--law cosmology driven by 
an exponential potential in the low--energy limit. 
Inflation proceeds indefinitely into the future if $C^2 <2$
and the expansion decelerates for $C^2 >2$. However, 
at early times, the asymptotic behaviour is $a \propto 
t^{1/C^2}$, and inflation proceeds for a finite time for 
$C^2<1$. In this limit, the potential is of the form 
$V \propto \phi^{-2}$, where 
the constant of proportionality
determines the power of the expansion. 

At late times the function $y$ given in Eq. (\ref{ansatz1}) 
asymptotes to an exponential form. It is therefore of interest 
to consider a second ansatz
\begin{equation}
\label{ansatz2}
y = \exp \left( -\sqrt{2\pi} C \phi /m_4 \right)  ,  
\end{equation}
that is valid for all time, 
where we assume implicitly that $\phi >0$ 
and that $C^2<6$. Integrating Eqs. 
(\ref{fprime}) and (\ref{fscale}) implies that 
\begin{eqnarray}
\label{time1}
\phi = \frac{m_4}{\sqrt{2\pi}C} 
\ln \left( T +\sqrt{T^2-1} \right) \\
\label{time2}
a= \left( T +\sqrt{T^2 -1} \right)^{2/C^2}   , 
\end{eqnarray}
where we have introduced a rescaled time variable
\begin{equation}
T \equiv \left( \frac{\pi \lambda C^4}{3 m_4^2} 
\right)^{1/2} 
\left( t-t_0 \right)
\end{equation}
and the origin of time corresponds to 
$T=1$. 
The potential of the scalar field is deduced by substituting 
Eq. (\ref{ansatz2}) into Eq. (\ref{fpotential}): 
\begin{equation}
\label{secondpotential}
V= \frac{2\lambda y^2}{1-y^2} \left[ 
1-\frac{C^2}{6} \frac{1}{1-y^2} \right]   . 
\end{equation}
The potential is negative for 
$y^2 > (6-C^2)/6$, has a single maximum 
located at 
$y^2=(6-C^2)/(6+C^2)$ 
and exponentially 
decays to zero from above as $y^2 \rightarrow 
0$. 
Although the potential is negative to 
the right of the maximum, the 
solution exists for all $\phi >0$ because the 
initial magnitude of the field's 
kinetic energy 
is sufficiently large for
it to move over the maximum and 
reach $\phi \rightarrow +\infty$. 

We now consider the formulation 
of the cosmological brane equations summarized in 
Eqs. (\ref{definitionb})--(\ref{btime}). 
One ansatz that can be invoked is 
\begin{equation}
\label{bansatz}
b \equiv \left( \frac{3\pi A^2}{\lambda m^2_4} \right)^{1/2}
\phi   , 
\end{equation}
where $A$ is a constant. From Eq. (\ref{bkinetic}), this 
represents a model 
where the kinetic energy of the scalar field is a constant
for all time: 
\begin{equation}
\phi =\phi_0 -A \left( t-t_0 \right)    . 
\end{equation}
The scale factor and potential of the field are 
readily deduced from Eqs. (\ref{bscale}) and 
(\ref{bpotential}), respectively: 
\begin{eqnarray}
a = \exp \left[ -\frac{ \lambda}{3A^2} \cosh 
\left( \sqrt{\frac{12\pi A^2}{\lambda m^2_4}} \phi 
\right) \right] \\
V = 2 \lambda \sinh^2 \left( \sqrt{\frac{3\pi A^2}{\lambda m^2_4}} 
\phi \right) -\frac{A^2}{2}  . 
\end{eqnarray}
Early times correspond to the 
limit $\phi \gg \lambda^{1/2} m_4/A$, where the 
potential has an asymptotically exponential form
and  
inflation may proceed for 
a wide range of parameter space. 
Inflation ends within a 
finite time, 
however, and the field eventually 
falls into its minimum at 
$V_{\rm min} =-A^2 /2$. The expansion of the 
universe is 
then reversed and the subsequent collapse enables the 
field to move up the other side of its potential.

Another solvable model is defined by 
\begin{equation}
\label{bansatz2}
b \equiv p_2 \phi^2  , 
\end{equation}
where $p_2$ is an arbitrary constant. Substituting 
Eq. (\ref{bansatz2}) into Eq. (\ref{bpotential}) 
implies that the potential is given by
\begin{equation}
\label{masspot}
V= 2 \lambda \sinh^2 \left( p_2 \phi^2 \right) - 
\frac{2\lambda p^2_2m^2_4}{3\pi} \phi^2   . 
\end{equation}
This potential has a single maximum at 
$V(\phi =0 )=0$ and two minima at 
\begin{equation}
\sinh 
(2p_2\phi^2) = \frac{p_2m^2_4}{3\pi}  .
\end{equation}
Integration of Eq. (\ref{bkinetic}) 
implies that the time--dependence of the 
scalar field is given by
\begin{equation}
\label{phiunexpected}
\phi = \phi_0 \exp \left[ 
- \left( \frac{4\lambda p_2^2 m_4^2}{3\pi}
\right)^{1/2} \left( t-t_0 \right) \right]   . 
\end{equation}
Finally, the growth in the 
scale factor can be expressed as 
a power series by substituting Eq. (\ref{bansatz2})
into Eq. (\ref{bscale}) and integrating: 
\begin{equation}
\label{scaleseries}
a = \exp \left[ - \frac{\pi}{2p_2 m_4^2} \sum_{n=0}^{\infty}
\frac{1}{(2n+1)(2n+1)!} \left( 2p_2 \phi^2 \right)^{2n+1}
\right]    . 
\end{equation}

Before concluding, we develop in the next Section 
an  algorithm that generates a new braneworld cosmology 
from a known solution such as those presented above.

\section{A Third Algorithm}

\setcounter{equation}{0}

Since 
$b (t)$ is a monotonically decreasing 
function, Eq. (\ref{bkinetic})
may be rewritten as
\begin{equation}
\label{rewritebkinetic}
\dot{\phi}^2 = - \left( \frac{\lambda m^2_4}{3\pi} \right)^{1/2}  
\dot{b} . 
\end{equation}
Defining a new function
\begin{equation}
\label{cfunction}
c \equiv \exp \left[ -\int^t dt' 
b(t') \right]
\end{equation}
and a new time parameter
\begin{equation}
\label{eta}
\eta  \equiv \int^t dt' c(t')
\end{equation}
implies that Eq. (\ref{rewritebkinetic}) 
may be then expressed in the form 
of a one--dimensional Helmholtz equation: 
\begin{equation}
\label{helmholtz}
\left[ \frac{d^2}{d\eta^2}  - 
U(\eta ) \right] c =0   , 
\end{equation}
where the effective potential is 
uniquely determined 
by the kinetic energy of the scalar field: 
\begin{equation}
\label{effectiveequation}
U (\eta ) \equiv \left( \frac{3\pi}{\lambda m^2_4} 
\right)^{1/2} \left( \frac{d \phi}{d\eta} 
\right)^2   . 
\end{equation}

The importance of Eq. (\ref{helmholtz}) 
is that if a particular solution, $c_1 (\eta)$, 
is known, the {\em general} solution 
can be written in terms of a single quadrature 
with respect to this solution \cite{nieto}: 
\begin{equation}
\label{generalquad}
c = c_1 \left( \kappa +\int^{\eta} \frac{dz}{c_1^2 (z)}
\right)   , 
\end{equation}
where $\kappa$ is an arbitrary constant. 
It is this feature that 
form the basis of the algorithm. 
Suppose a  
particular solution 
to the braneworld field equations has already been 
found, i.e., that $\{ b (t) , \phi( t) \}$ 
are known. 
The function, $c_1 (\eta )$, can 
in principle 
be evaluated from Eqs. (\ref{cfunction}) and (\ref{eta}). 
Eq. (\ref{generalquad}) then 
yields the new solution for $c (\eta)$
and, hence, $b(\eta )$ from the definition 
(\ref{cfunction}). The form of $\phi (\eta )$ 
is identical in both solutions, but the potential 
of the scalar field is different in the 
new solution. It is given by 
\begin{equation}
\label{generalpotential1}
V [ \eta (\phi ) ] = 2\lambda {\rm sinh}^2 
\left[ b(\eta ) \right] - \frac{1}{2} c^2(\eta )
\left( \frac{d \phi}{d \eta} \right)^2 , 
\end{equation}
or, equivalently, by
\begin{equation}
\label{generalpotential2}
V [ \eta (\phi) ] =2 \lambda \left[ {\rm sinh}
\left( \frac{d c}{d \eta} \right) \right]^2 - \left( 
\frac{\lambda m^2_4}{12 \pi} \right)^{1/2} 
c \frac{d^2c}{d \eta^2}  . 
\end{equation} 
The form of $V(\phi)$ follows from Eqs. 
(\ref{generalpotential1}) and (\ref{generalpotential2}) 
and the scale factor is deduced 
as before from 
Eq. (\ref{bscale}).

In effect, each particular braneworld cosmology is 
twinned with  a second solution. In practice, 
it may not always be possible to 
evaluate the integrals (\ref{cfunction}) 
and (\ref{eta}), but 
it would be interesting to explore models of this nature 
further. We also remark that 
in the low--energy limit, $c$ 
varies as a power of the scale factor 
of the brane. 
Thus, the above discussion 
is also relevant to 
standard scalar field cosmology.  

\section{Conclusions}

\setcounter{equation}{0}

In conclusion, 
we have presented algorithms for 
solving the braneworld Friedmann equation 
for a single, 
self--interacting scalar field confined to a
brane with positive tensione3 embedded in 
five--dimensional, vacuum Einstein gravity.  
The formalism does not assume the slow--roll 
approximation and
is valid  
during inflation when the field is 
monotonically rolling 
down its potential. It does not apply 
if the field is oscillating about a minimum. 
A number of new exact solutions were found, 
including   
the scalar field model that corresponds to 
a perfect fluid. The algorithms
may be viewed as generalizations of the 
Hamilton--Jacobi formalism that has played a central role 
in analyses of the standard chaotic inflationary scenario 
\cite{hj}. 
In the latter case, the Hamilton--Jacobi 
formalism provides the necessary framework 
for establishing the precise 
correspondence between 
the potential 
of the scalar  field and the 
scalar and tensor perturbation spectra that are generated 
during inflation \cite{infrev1}.  
It would be interesting to employ 
the techniques developed above to 
establish the equivalent correspondence in the braneworld 
scenario.

\acknowledgments

RMH is supported by a Particle Physics 
and Astronomy Research Council (PPARC) postgraduate 
studentship.  JEL is supported by the Royal Society.
We thank David Wands for numerous helpful discussions.

\end{document}